\documentclass[11pt,a4paper]{article}

\usepackage[margin=2cm]{geometry}
\usepackage[utf8]{inputenc}
\usepackage{amsmath,amssymb,amsfonts,amsthm}
\usepackage{graphicx}
\usepackage[hidelinks]{hyperref}
\usepackage{authblk}
\usepackage[authoryear]{natbib}

\title{Supershear-subshear-supershear rupture sequence during the 2025 Mandalay Earthquake in Myanmar}

\date{2025-07-30}

\author[1]{Shiro Hirano\thanks{\texttt{hirano@hirosaki-u.ac.jp}}}
\author[1]{Ryosuke Doke}
\author[1]{Takuto Maeda}
\affil[1]{Hirosaki University, 3, Bunkyo, Hirosaki city, Aomori, 036-8561, Japan}

\begin{document}

\maketitle

\begin{abstract}
        We investigated the rupture dynamics of the 2025 $M_w$7.7 Mandalay, Myanmar earthquake, using a video recording of surface rupture, strong motion recordings, waveform simulation, and satellite imagery. Our assessment, based on the S-wave observation in the video and rupture arrival time at a seismic station 246 km south of the hypocenter, suggests that rupture decelerated to subshear speeds ($\sim$3 km/s) from initial supershear propagation ($\sim$6 km/s) before reaching the camera location. This deceleration is also supported by comparison between the fault-normal acceleration patterns seen in the video and that simulated by kinematic rupture modeling. Additionally, satellite imagery indicated a local minimum in slip (2--3 m) approximately 40--60 km south of the epicenter, suggesting a region of reduced stress drop that likely caused the temporary deceleration. Beyond this point, the rupture appears to have re-established supershear propagation.
\end{abstract}

The published version is available at \url{https://doi.org/10.26443/seismica.v4i2.1785}

\section{Introduction}

Describing coseismic rupture, surface displacement, and deformation near a fault remains a critical challenge for both seismology and earthquake engineering.
Strong motion records within a few hundred meters of the fault have been obtained from events such as the 2000 Western Tottori earthquake \citep{Fukuyama07} and the 2023 Kahramanmaraş earthquake \citep{METU23}. However, such observational opportunities are limited due to the cost of installing instruments and the difficulty in predicting rupture locations, which would support deployment of stations well in advance.
Therefore, it is important not to rely only on seismic recordings, but to use whatever available data to study the rupture characteristics. 

At 06:20 UTC on March 28, 2025, an $M_w$7.7 earthquake occurred along the Sagaing fault, which traverses Myanmar from north to south (Fig.S3).
The epicenter was located at $22.001^\circ$N, $95.925^\circ$E, with a focal depth of 10 km \citep{USGS25}.
The event involved the rupture of a near-vertical strike-slip fault spanning approximately 450 km\citep{USGS25,Inoue25}.
Severe damage was reported in Myanmar and Thailand, with numerous instances of surface rupture, especially observed in areas along the fault trace.

On May 11, 2025, a video purportedly capturing coseismic surface rupture due to the mainshock was made available online (for details, see Data and code availability section, and Fig.S3 for the location).
The footage was taken approximately 124 km south of the epicenter.
Based on the posted latitude and longitude information ($20.8821^\circ$N, $96.0353^\circ$E), the camera, hereafter referred to as CCTV, was located within a megawatt solar power plant adjacent to the northern part of Thazi town.

The shadows seen in the video suggested that the camera was facing southwest \citep{Latour25}.
The timestamps in the upper right corner of the frames appear to be about five minutes behind the local standard time when compared to the USGS earthquake origin time \citep{USGS25}.
The timestamps are in \texttt{hh:mm:ss} format. Hereinafter, the seconds component (\texttt{ss}) is defined as $T$.
The footage indicated that subtle shaking had begun by $T=$30, which intensified after $T=$33.
Between $T=$35 and $T=$37, the footage clearly shows right-lateral strike-slip displacement of the ground beyond the fence on the right side of the video frames and the foreground gate.
This recording is likely the world's first direct observation of surface fault rupture during an earthquake. It provides direct information about the rupture at a location where no seismographs were installed.

An initial analysis of the video revealed a pulse-like slip velocity function that completed within approximately two seconds \citep[see Supplementary Material for the method]{Hirano25}.
Although the dimension of the initial result was pixels/s, it has since been refined through the analysis of numerous additional points within the video, indicating that the maximum slip velocity reached 3 m/s \citep{Latour25,Kearse25}, or 4.5 m/s considering field observations \citep{Gao25}.

For the Myanmar earthquake, information on near-field ground motion (e.g., within 10 km from the fault trace) is likely limited to the video analysis described above and waveform data from the GEOFON station in Naypyidaw (NPW) \citep{Lai25}.
This lone station is located approximately 246 km south of the epicenter and about 2.6 km west of the fault.
Consequently, it is crucial to leverage all potential data and explore diverse avenues to understand the source behavior of this earthquake.

The rupture primarily propagated southward from the hypocenter.
An inversion result suggests the propagation of a supershear rupture around the hypocenter \citep{Inoue25}.
Furthermore, in the time-corrected fault-parallel displacement waveforms \citep{Lai25}, a rapid displacement at NPW, likely associated with the passage of the rupture front, began approximately 50 s after rupture initiation, with no clear S-wave priorly observed.
The average rupture velocity during this interval is $\frac{246}{50}=4.92$ km/s.
Even based solely on this record, it is strongly inferred that a significant portion of the rupture propagated at supershear velocities.

Conversely, at the CCTV location, which is closer to the epicenter, the rupture might propagate at subshear rather than supershear speeds.
This is supported by \cite{Latour25}, who note that the S-wave appears to arrive 1.8 seconds earlier than the rupture front at this site.
While the video's timestamp is inaccurate, which makes the exact arrival times of the S-wave and rupture at the camera location unknown, if a substantial portion of the rupture propagated subshear from the initiation point to the CCTV location, it would contradict the inversion results \citep{Inoue25}.
Therefore, a change in the rupture propagation velocity would be consistent.

In this paper, we estimate the history of ground motion and rupture propagation velocity using information from both the video and strong motion records.
Specifically, we demonstrate that the fault rupture decelerated to subshear speeds just before reaching the CCTV location, although it generally propagated at supershear speed.

\section{Methods \& Results}

\subsection{Restriction of rupture front}

Herein, the origin time of the earthquake, 2025-03-28 06:20:52 UTC \citep{USGS25}, is defined as $t=0$ s.
As previously stated, the rupture arrival time at the NPW station is estimated to be $t_N=50$ s.
We define $r$ (km) as the epicentral distance, with the CCTV location at $r_C=124$ km and the NPW station at $r_N=246$ km.
Assuming a P-wave velocity of $V_P=6.0$ km/s, and given that the rupture cannot propagate faster than this velocity, the rupture front at any given location where $r \le r_N$ must arrive within the time window defined by $V_P^{-1} r \le t \le V_P^{-1} r+t_N-V_P^{-1} r_N$.
The lower bound of this inequality signifies that rupture cannot occur before the arrival of the P-wave radiated from the hypocenter.
The upper bound accounts for the allowable delay time for the rupture to reach NPW at $t=t_N$.
If this inequality is violated at any $r(<r_N)$, the rupture cannot arrive at $r_N$ at time $t=t_N$, even if its subsequent propagation occurs at the P-wave velocity.

\begin{figure}
\centering
\includegraphics[width=\linewidth]{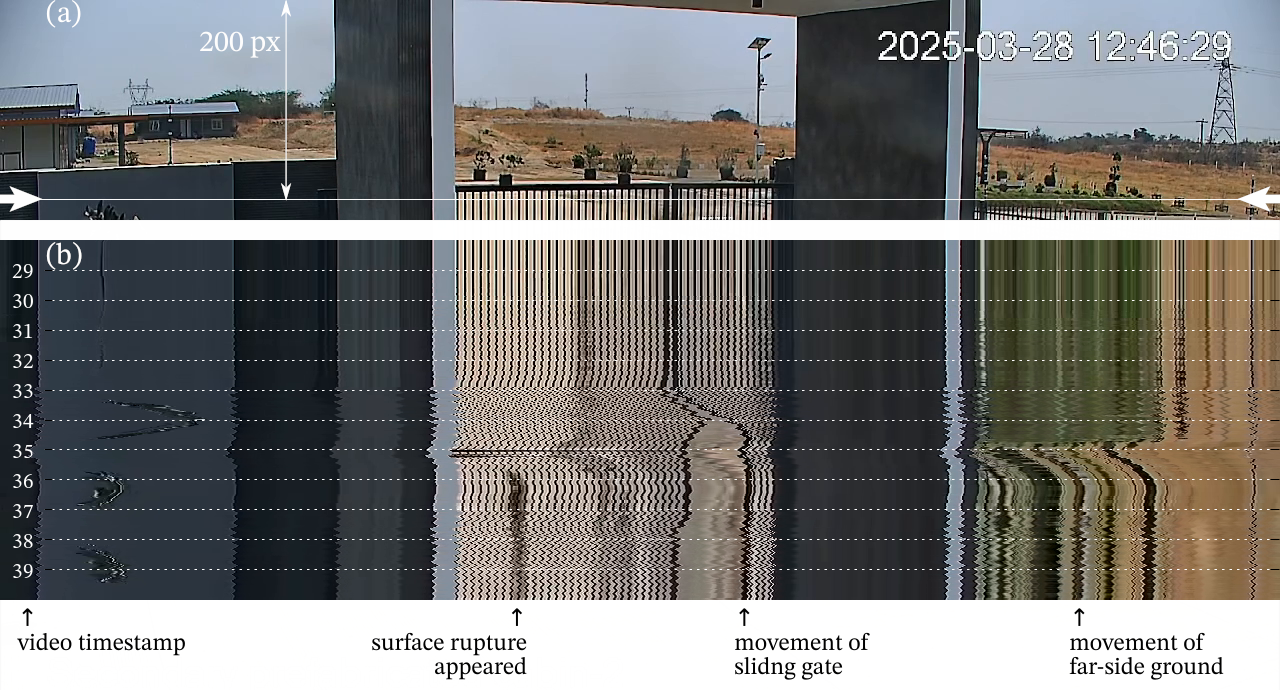} 
\caption{
    (a) The upper part of the original video frame. The white line between the two horizontal arrows at 200 pixels from the top edge indicates the extracted line in the following process.
    (b) A stacked image obtained by extracting a 1-pixel-height slice along the horizontal line in (a) and aligning these slices vertically.
The tics and values within (b) correspond to the timestamp in seconds displayed in the upper right corner of the original video; the timestamp appears to be approximately five minutes behind the actual local time when compared to \cite{USGS25}.}
\label{lines_from_video}
\end{figure}

Next, we focus on the observation of a clear S-wave at the CCTV location.
To investigate its duration, we analyzed the video footage, which has a resolution of 1280$\times$720 pixels and a frame rate of 30 fps.
From each frame of the video, we extracted a 1-pixel-height horizontal line at 200 pixels from the top edge, spanning 1280 pixels in width.
These extracted lines were then aligned vertically to construct a stacked image (Fig.\ref{lines_from_video}).

Focusing on the movement of the sliding gate depicted in Fig.\ref{lines_from_video}, the subtle shaking started no later than $T=30$ s is likely attributed to the P-waves originating from the hypocenter.
We have to note that the exact onset cannot be determined from the image analysis and the P-wave could have arrived before $T=30$ s.
This is because the initial phase amplitude is typically low and we cannot estimate the sensitivity of the image to ground shaking, which inherently depends on the rigidity of the structure holding the camera.
Subsequently, intense shaking commences at $T\sim33$ s, which is interpreted as the arrival of S-waves. 
Following this, observations of the right side of the video frames indicate that fault slip initiated around $T=35$ s, with the slip completing within approximately two seconds.

\begin{figure}
\centering
\includegraphics[width=86mm]{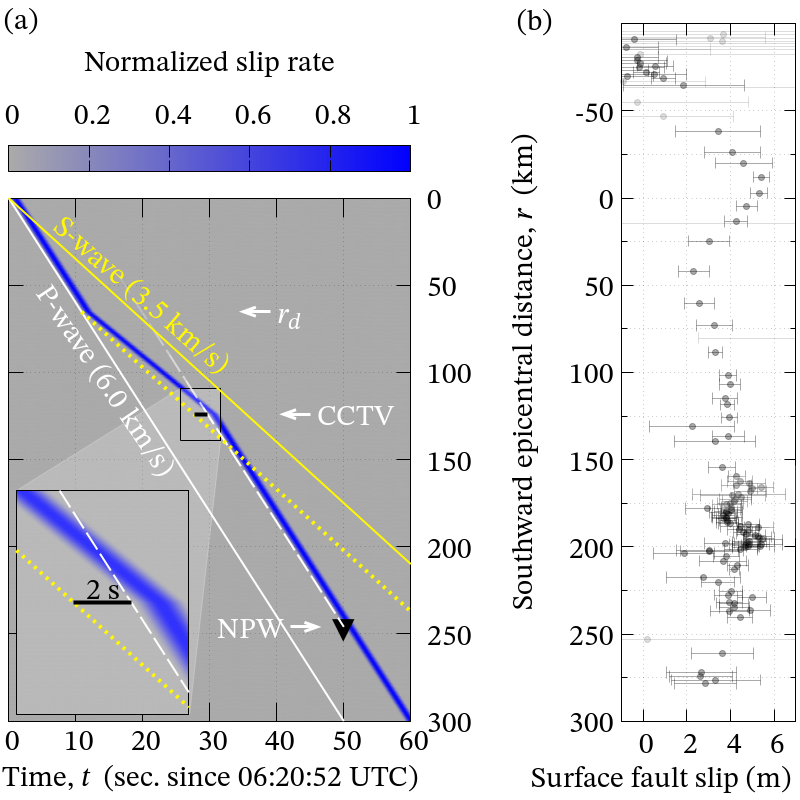} 
\caption{(a) A potential rupture scenario illustrating supershear-subshear-supershear propagation from the epicenter to the NPW station (triangle, hypocentral distance $r=$ 246 km). 
The gray-to-blue color contour represents the pulse-like slip velocity with a duration of two seconds.
The solid white line represents the direct P-wave radiated from the hypocenter.
The dashed white line indicates the condition under which the rupture reaches the NPW station at $t=50$ s.
If the rupture front crosses to the right of this line, it cannot reach the NPW station by $t=50$ s.
Solid and dotted yellow lines indicate propagation velocity of 3.5 km/s through the hypocenter and the deceleration point, respectively.
The inset shows a zoomed plot where the horizontal black bar indicates the two-second duration of the S-wave before the rupture onset observed at the CCTV site ($r=$ 124 km).
(b) Slip distribution along the surface fault trace derived from Sentinel-2 imagery. 
The ordinates are common to both (a) and (b).
See Supplementary Material for algebra to obtain the deceleration point ($r_d =$ 65--99 km) and subshear rupture velocity (2.7--3.1 km/s) in (a) and an image analysis to calculate the slip in (b).
}
\label{rupture_slip}
\end{figure}

Given the earthquake origin time, the arrival times of waves and rupture at NPW, and the duration of S-wave arrival and slip at CCTV, the hypothetical rupture propagation history is estimated as depicted in Fig.\ref{rupture_slip}a.
The rupture front must always remain between the two white lines in Fig.\ref{rupture_slip}a.
Despite the rupture propagation velocity being supershear near the rupture initiation point \citep{Inoue25}, the observation of an approximately two-second S-wave (indicated by the horizontal black bar in Fig.\ref{rupture_slip}a) before the rupture front arrival at the CCTV location suggests that the rupture must have decelerated to about 2.7--3.1 km/s at 65--99 km from the hypocenter as calculated in Supplementary Material.

\subsection{Fault-normal acceleration prior to the rupture}

The ground motion observed at the CCTV site immediately preceding the onset of rupture suggests subshear rupture propagation. A key observation for that is the arrival of S-wave prior to the rupture \citep{Latour25}.
The other one is the rightward displacement of the sliding gate during the approximately two-second interval between the S-wave arrival and the initiation of fault slip as seen in Fig.\ref{lines_from_video}. This indicates that the ground experienced substantial acceleration towards the east (left side of the video frames), causing the sliding gate to be displaced apparently westward due to inertia.

Given that this is a right-lateral strike-slip fault, the rupture is expected to propagate from the north (right foreground of the video frames) to the south (left background). Generally, as such slip propagates, the ground ahead of the rupture front moves towards the left of the video frames for subshear rupture and towards the right of the video frames for supershear rupture, as shown by 2-D steady-state pulse solutions \citep{Dunham05} and 3-D dynamic rupture simulations \citep{Abdelmeguid24}. However, these models do not account for abrupt rupture deceleration directly before the observation point. Therefore, to fully understand this event, it would be beneficial to interpret the results of kinematic simulations where rupture propagation velocity matches observations from the Myanmar event.

\begin{figure}
\centering
\includegraphics[width=86mm]{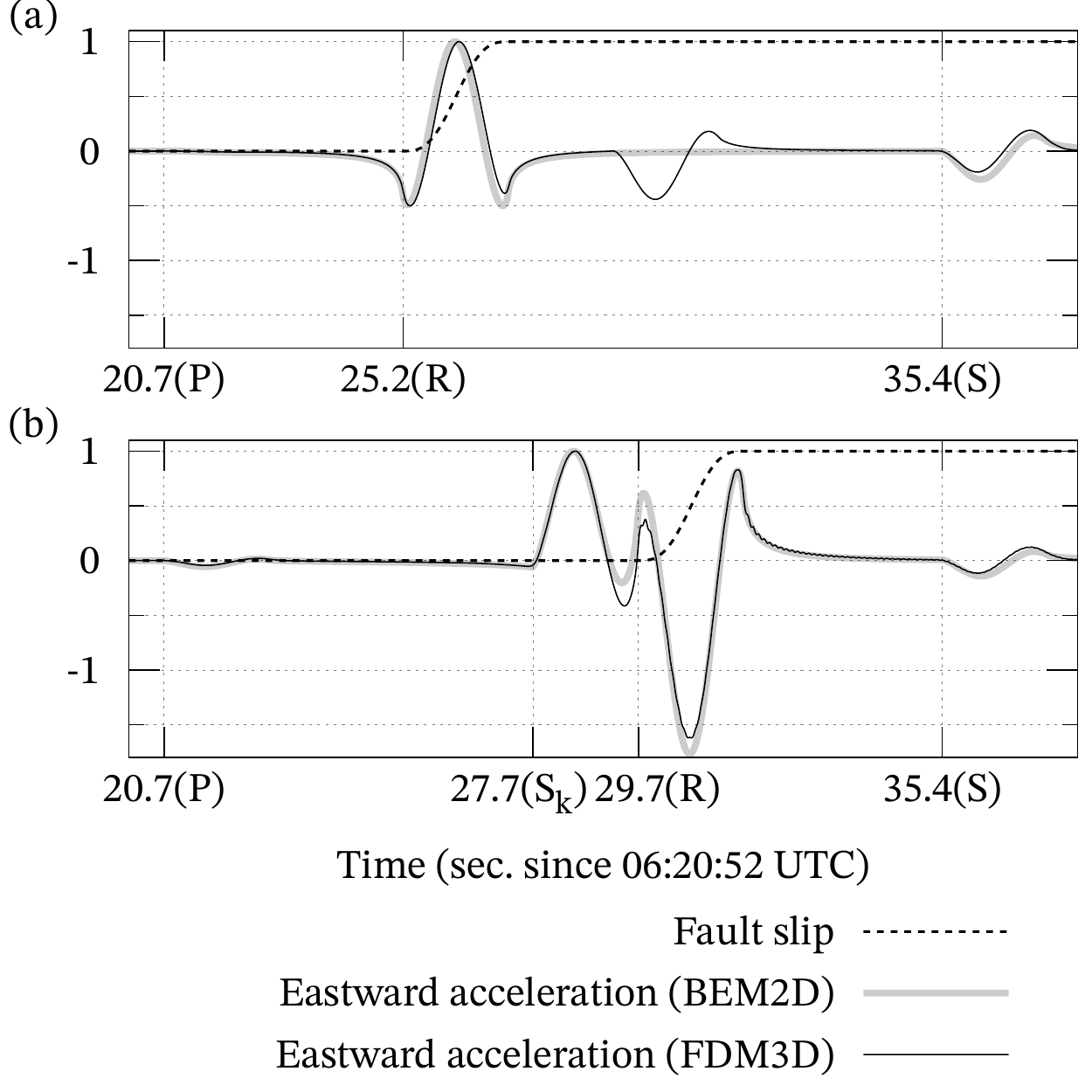} 
\caption{The assumed slip history (dashed black) and the calculated fault-normal acceleration (solid gray for 2-D and solid black for 3-D) as a two-second duration slip pulse propagates past the CCTV location.
The rupture scenarios are (a) the supershear rupture propagating with uniform velocity of 4.92 km/s and (b) the supershear-subshear-supershear rupture as in Fig.\ref{rupture_slip}a.
The labels P, S${}_\textrm{k}$, R, and S indicate arrivals of P-wave from the hypocenter, S-wave from the rupture deceleration point, the rupture front, and S-wave from the hypocenter, respectively.
For the solid lines, positive values indicate eastward acceleration.
All units are arbitrary.}
\label{Calculations}
\end{figure}

We calculated the fault-normal acceleration at a distance of $r=124$ km from the epicenter with 2-D and 3-D models. The 2-D modeling was achieved by employing the integral kernel derived for 2-D velocity fields within the framework of the Boundary Integral Equation Method \citep{Tada00}. 
The 3-D modeling with a free surface and a planar strike-slip fault down to 20 km depth was done using OpenSWPC, an open-source code utilizing a finite difference method \citep{Maeda17}.
The methodology on the 2-D and 3-D calculations are detailed in Supplementary Material.
For the fault slip time history, we assumed two distinct slip pulses: uniform supershear rupture with the propagation speed of 4.92 km/s as calculated in the introduction and supershear-subshear-supershear rupture sequence as illustrated in Fig.\ref{rupture_slip}a. Both scenarios involve a slip velocity, which is given by
$
\dot{D}(t) = 
\sin^2{\dfrac{\pi(t-t_r)}{2}} H(t-t_r) H(t_r + 2 - t)
$
with a duration of two seconds, where $H(\cdot)$ is the Heaviside function.
The rupture initiation time, $t_r$, is dependent on the epicentral distance, $r$, and follows the scenarios. 
While the actual duration of fault slip can exceed two seconds at various locations, with some areas peaking more than 10 seconds after rupture initiation \citep{Inoue25}, the impact of these longer durations would appear in the right side of the dashed white line in Fig.\ref{rupture_slip}a.
Therefore, given that our current analysis focuses only on the period immediately surrounding the rupture's arrival, these longer slip durations are disregarded.

Our modeling reveals distinct fault-normal acceleration patterns depending on the assumed rupture propagation. In the case of constant supershear rupture (Fig.\ref{Calculations}a), only a gradually increasing westward acceleration is observed prior to the onset of slip.
Conversely, an abrupt change in rupture propagation velocity (Fig.\ref{Calculations}b) leads to the generation of a significant eastward acceleration commencing two seconds prior to slip initiation. 
This finding demonstrates robustness, having been reproduced by both 2-D and 3-D simulations. 
The discrepancies observed after $t \sim 29$ s are attributed to the finite fault extent in the 3-D case, where strong radiation also appears from the bottom edge of the fault.
Unlike the constant velocity scenario, Fig.\ref{Calculations}b also shows discernible ground motion caused by P-waves, which is consistent with the observations from the video analysis.
Therefore, these results suggest that the rupture decelerated prior to reaching the CCTV location.

\section{Discussion}

To investigate the reason for the abrupt deceleration of the rupture, we calculated the surface fault slip distribution (Fig.\ref{rupture_slip}b) using satellite imagery.
North-south surface displacements obtained at 80-m resolution reveal that slip occurred precisely along the known fault trace (Fig.S3a). 
By taking their difference between the east and west, we can derive the surface slip distribution (see Supplementary Material for details).
The results show that the slip reached 5 m near the epicenter. 
However, it exhibits a local minimum of 2--3 m at 40--60 km south of the epicenter. 
Further south, a widespread average slip of 4 m, without a broad region of reduction, is distributed over an epicentral distance of approximately 100--250 km.

Final slip amount correlates with the static stress drop.
Mathematically, the slip distribution $D(\boldsymbol{k})$ and stress drop distribution $\tau(\boldsymbol{k})$ on a planar fault are related as $D(\boldsymbol{k}) \sim - \dfrac{2}{\mu |\boldsymbol{k}|} \tau(\boldsymbol{k})$, where $\boldsymbol{k}$ is the 2-D wavenumber vector \citep{Andrews80}.
This indicates that $D$ is obtained by reducing the shorter wavelength (i.e., larger $|\boldsymbol{k}|$) components of $\tau$.
This relationship, resulting from numerical simulations with heterogeneous stress drop, is also evident in Figs. 3 and 4 of \cite{Andrews11}.
Considering this, a large static stress drop can be inferred near the epicenter of the Myanmar earthquake.
Conversely, a region with a small stress drop likely existed, centered 40--60 km from the epicenter ($\sim21.5^\circ$N). 

In general, local rupture acceleration stems from the energy release rate, which scales with the square of the stress intensity factor. The stress intensity factor itself scales with stress drop (Freund 1990).
This indicates a theoretical local correlation between stress drop and rupture velocity.
Therefore, although the rupture accelerated to supershear speed near the epicenter due to the large stress drop, energy release gradually weakened around the 40--60 kilometer point, preventing it from maintaining supershear propagation.
Theoretically, rupture propagation between the Rayleigh wave speed and $\sqrt{2}$ times the S-wave speed is unstable because the stress singularity is weaker than the square root \citep[section 4.3.4]{Freund1990}.
Consequently, the energy release rate decreases as the rupture decelerates within this speed range.
Hence, the rupture might immediately revert to subshear regime due to the positive feedback between rupture deceleration and energy release rate reduction.
Subsequently, beyond the 100-kilometer point, increased stress drop and energy release may have led to the re-establishment of supershear rupture.

For the location $\sim21.5^\circ$N, \cite{Inoue25} revealed the local minimum of the released potency density and the variation of dipping angle of the fault by their potency density tensor inversion method that enables to image not only slip amount but also strike/dip/rake angles on each point along the fault plane.
Also, some investigations have revealed that the location $\sim21.5^\circ$N is a fault segment boundary characterized by the edge of a seismic gap and a discontinuity in the fault dip angle \citep{Hurukawa11,THAZINHTETTIN22}.
These seismological and geological results suggest that there exists a segment boundary that may behave as a barrier and reduce the slip at $\sim21.5^\circ$N.
Our model may thus provide a case study of the interaction between a dramatic change in rupture speed and a geological barrier.

\section{Conclusions}

This study provides unprecedented insights into the rupture characteristics of the 2025 Mandalay Earthquake ($M_w$7.7) along the Sagaing Fault. 
Through the integrated analysis of a unique coseismic video recording, seismic data, kinematic simulation, and satellite imagery, we have revealed a dynamic rupture history.

Our findings indicate that the rupture initiated with supershear velocities near the hypocenter but experienced a notable deceleration to subshear speeds as it approached the CCTV observation site. This temporary slowdown is strongly supported by observed fault-normal acceleration patterns and aligns with the presence of a localized minimum in fault slip around 40--60 km south of the epicenter, suggesting a zone of reduced stress drop. Subsequently, beyond this point, the rupture appears to have re-established its supershear propagation.

This research demonstrates the immense value of direct observations of surface fault rupture, especially in environments where traditional seismic instrumentation is sparse. The detailed rupture dynamics unveiled here contribute significantly to our understanding of complex earthquake behaviors, particularly concerning the factors influencing rupture velocity changes and their implications for near-fault ground motion. Future work will aim to further constrain the precise timing of these velocity transitions and their correlation with geological and stress conditions along the fault.

\section*{acknowledgements}

This study was supported by ERI JURP 2025-S-B102. This study was conducted using the FUJITSU Supercomputer PRIMEHPC FX1000 and FUJITSU Server PRIMERGY GX2570 (Wisteria/BDEC-01) at the Information Technology Center, the University of Tokyo.
Thorough reviews by Dr. Kiran Kumar Thingbaijam and Dr. Tiegan Hobbs contributed to improving the manuscript.
The authors thank Dr. Ryo Okuwaki for their helpful comments.

\section*{Data and code availability}

        The video, credited to GP Energy Myanmar, was originally uploaded to Facebook (\url{https://www.facebook.com/htin.aung.33/videos/1041579804084512}) on May 11, 2025, and later reposted on YouTube (\url{https://www.youtube.com/watch?v=77ubC4bcgRM}), where the CCTV site's coordinates were first disclosed.
        The significant portion of the video is featured in CBC video news (\url{https://www.cbc.ca/player/play/video/9.6763872}).
        The authors believe that using the video in this manuscript is consistent with the principles of fair use under U.S. copyright law for the purposes of scholarship and research (\url{https://www.copyright.gov/fair-use/}).
        All links in this paragraph were last accessed on July 15, 2025.

        The source code for 2-D BEM calculation is available at \url{https://doi.org/10.17605/OSF.IO/MTR76 }.
        OpenSWPC, the finite difference code, for our 3-D calculation is available at \url{https://doi.org/10.5281/zenodo.16264099}.
        The surface displacement in Supplementary Material is based on Copernicus Sentinel data (2025) processed by Sentinel Hub (\url{https://dataspace.copernicus.eu/explore-data/data-collections/sentinel-data/sentinel-2}).

\section*{Competing interests}
The authors have no competing interests.
		
\bibliographystyle{apalike-ejor}
\bibliography{mybibfile}

\newpage

\renewcommand{\thesection}{S.\arabic{figure}}
\renewcommand{\theequation}{S.\arabic{equation}}

\appendix

\section{Supplementary Material: Methods}

\subsection{Image analysis to extract slip history from the video}

In the initial analysis of the video \citep{Hirano25}, the author executed the following steps as shown in Fig.\ref{VideoAnalysis}:
\begin{itemize}
    \item Step 1: A 100$\times$20 px region, starting 152 px from the top and 930 px from the left, in video frames was cropped from each video frame.
    \item Step 2: The cropped images were compressed vertically to 100$\times$1 px, to obtain a vertical average.
    \item Step 3: Edges in the 1-px-height images were enhanced.
    \item Step 4: The distance between the leftmost and rightmost white pixels was measured.
\end{itemize}

\subsection{Calculations to obtain the deceleration point and subshear rupture velocity} \label{Sec:Algebra}

The initial supershear rupture velocity $V_\text{sup}$ falls between $V_P$ (= 6.0 km/s) and $V_S$ (= 3.5 km/s).
At the CCTV site, from the video frames, the P-wave duration $\tau_P$ is, though uncertain, at least 3 seconds, while S-wave duration $\tau_S$ is approximately 2 seconds.
The S-wave front arrival time therein is $t_C^S := V_P^{-1} r_C + \tau_P$, and the equation of the front (gray dotted line in Fig.\ref{Algebra}a) is given by
\begin{align*}
r = V_S \left(t - t_C^S\right) + r_C.
\end{align*}
The intersection of this line and the initial rupture propagation represented by $r = V_\text{sup} t$ gives the deceleration point which is the origin of the observed S-wave.
The solution $r_d$ is calculated as:
\begin{align*}
r_d
= & \frac{r_C - V_S t_C^S}{1 - V_S/V_\text{sup}}.
\end{align*}
The subshear rupture velocity $V_\text{sub}$ from the deceleration point to CCTV is
\begin{align*}
    V_\text{sub}
    = & \frac{r_C - r_d}{t_C^S + \tau_S - r_d/V_\text{sup}},
\end{align*}
where $t_C^S + \tau_S$ and $r_d/V_\text{sup}$ in the denominator represent the rupture times at the CCTV and the deceleration point, respectively.
From the above, unknown parameters are $\tau_P$ and $V_\text{sup}$. 
Given that rupture reached $r = r_N$ at $t = t_N$, the following condition must be satisfied:
\begin{align}
    t_C^S + \tau_S
    \le & t_N - \frac{r_N - r_C}{V_P}. \label{BackPropagation}
\end{align}
where the right-hand side denotes the boundary between the white and top-right gray regions in Fig.\ref{Algebra}a.
Otherwise, the rupture cannot reach NPW at $t=t_N$, even if it propagates at $V_P$ after the CCTV site.

We can read $t_N = 48$ from the Fig.4 of \cite{Lai25}, who assumed that the origin time was 2025-03-28 06:20:54 UTC, though a rationale was not provided.
This origin time is 2 seconds behind the USGS origin time, suggesting that $t_N = 50$ s may be more appropriate if the USGS time is adopted.
If $V_\text{sup} = V_P = 6$ km/s, the deceleration point and subshear rupture velocity for the earliest ($\tau_P=3$ s) and latest (the equality in \eqref{BackPropagation} holds) cases are $(r_d, V_\text{sub}) = (98.8, 2.74)$ and $= (65.2, 3.13)$, respectively.
Fig.\ref{Algebra}b indicates that $V_\text{sup} \le 5$ is difficult to achieve, as $V_\text{sub}$ becomes excessively slow, particularly in the earliest case. Furthermore, $V_\text{sup} < \sqrt{2} V_S \sim 4.95$ is unstable, as explained in the Discussion section.
For numerical simulations, the latest case was employed because $r_d = 65.2$ is more comparable to the local minimum of the surface slip as in Discussion section.

\subsection{2-D kinematic rupture simulation by a boundary element method}

We calculated on-fault acceleration during the rupture propagation by 2-D boundary element method.
The numerical integration is performed by calculating equation (32) of \cite{Tada00}.
In their notation, the flat 2-D crack is along the $x_1$-axis.
We consider strike-slip faulting, where the slip direction and fault-normal direction are along the $x_1$- and $x_2$-directions, respectively.
Hence, our equation is 
\begin{align}
\dot{u}_2^{i,j,n} = \sum_{k,m} D_1^{k,m} \, \dot{K}_{1,u2}^{i,j,k,n,m}, \label{Tada2001eq32}.
\end{align}
Differentiating eq.\eqref{Tada2001eq32} with respect to time yields the acceleration waveform.

The slip-rate function is 
\begin{align}
D_1^{k,m} = \sin^2\left(\frac{\pi}{2}\left(t_m - t_r^k\right)\right) \, H\left(t_m - t_r^k\right) \, H\left(t_r^k+2-t_m\right), \label{SlipRateFunction}
\end{align}
where $H(\cdot)$ is the Heaviside function, and $t_m$ is the $m$-th time collocation point.
$t_r^k$ is the rupture arrival time at $x = x_k$, where $x_k$ is the $k$-th discretized fault segment edge.
For Fig.3a and 3b in the main text, we employed the following two scenarios, respectively, with some parameters obtained in \ref{Sec:Algebra}:
\begin{itemize}
\item If the rupture velocity is 4.92 km/s (constant) everywhere, then $t_r^k = \dfrac{x_k}{4.92}$.
\item If rupture velocity is 3.13 km/s in 65.2 km $< x <$ 124 km and is 6 km/s otherwise, then $t_r^k = \begin{cases}
\frac{1}{6}x_k & (0 \le x_k \le 65.2) \\
\frac{65.2}{6} + \frac{1}{3}(x_k - 65.2) & (65.2 < x_k < 124) \\
\frac{65.2}{6} + \frac{1}{3}(124 - 65.2) + \frac{1}{6}(x_k - 124) & (124 \le x_k) \\
\end{cases}$.
\end{itemize}

\subsection{3-D kinematic rupture simulation by a finite difference method}

We numerically solved the equations of motion and constitutive relations for a generalized Zener body using the 3-D parallel finite difference code OpenSWPC \citep{Maeda17, Maeda25}. A 3-D medium with dimensions of 500 km (north-south) $\times$ 200 km (east-west) $\times$ 100 km (vertical) was discretized using a uniform grid with 100-meter spacing in each direction. The medium was assumed to be a homogeneous, viscoelastic half-space with a mass density of 2.7 g/cm${}^3$, P-wave velocity of 6.0 km/s, S-wave velocity of 3.5 km/s, and intrinsic attenuation factors of Qp = 600 and Qs = 300. The ground surface was defined at 5 km from the top of the domain, and traction-free boundary conditions \citep{Nakamura12} were applied there. The earthquake source was represented as a collection of discrete point sources. A rectangular region in the central part of the model, extending 300 km in the north-south direction and 20 km in depth from the ground surface, was designated as the fault plane. Stress changes associated with fault slip \citep{Coutant95, Pitarka99} were applied to the numerical grid points corresponding to the fault for the duration of the rise time starting from the rupture onset time. The rupture onset time at each point was computed according to the setup described in the main text, with rupture initiation assumed at the northern end of the fault. To suppress artificial reflections from the model boundaries, the ADE-CFS perfectly matched layer scheme \citep{Zhang10} with a thickness of 20 grid points was applied along the outer edges of the computational domain. The model was partitioned into 32 $\times$ 32 segments in the horizontal directions, and the simulation was executed on 256 CPUs. Four MPI processes were launched per CPU, with each process utilizing one-fourth of the CPU's cores via OpenMP threading. The computation of 60,000 time steps with a time increment of 0.005 s took approximately 30 minutes.

\subsection{Estimation of Surface fault slip}

We estimated coseismic surface fault slip by pixel-offset (optical correlation) analysis of pre- and post-seismic Sentinel-2 satellite imagery. Here, we used 10 m-resolved Band-3 images of the L2A product, with a central wavelength of 560 nm, obtained on March 20th and 30th, 2025.

For the pixel-offset analysis, we used geoCosiCorr3D  \citep{Aati22a,Aati22b}. Here, we employed 32 pixels (320 m) as a window size for correlating the images, and estimated northward and eastward displacement for every eight steps (80 m). Estimated surface displacements were mapped in Fig.\ref{Sentinel2}.

The distribution of surface fault slip (Fig.2b in the main text) was estimated from the discrepancies in northward components between both sides of the surface fault trace, assuming that this component corresponds to the fault displacements, as the fault strike is almost north-south, and the east-west component was negligible. 
Considering the window size of the pixel-offset analysis, we excluded 1 pixel on the fault trace and 2 pixels (160 m) on each side from the fault in the analysis results.
We then calculated the average displacements ranging from 3 to 12 pixels (800 m) on each side of the fault on the east-west cross-sections.


\renewcommand{\thefigure}{S.\arabic{figure}}
\setcounter{figure}{0} 

\begin{figure}[h!]
\centering
\includegraphics[width=\linewidth]{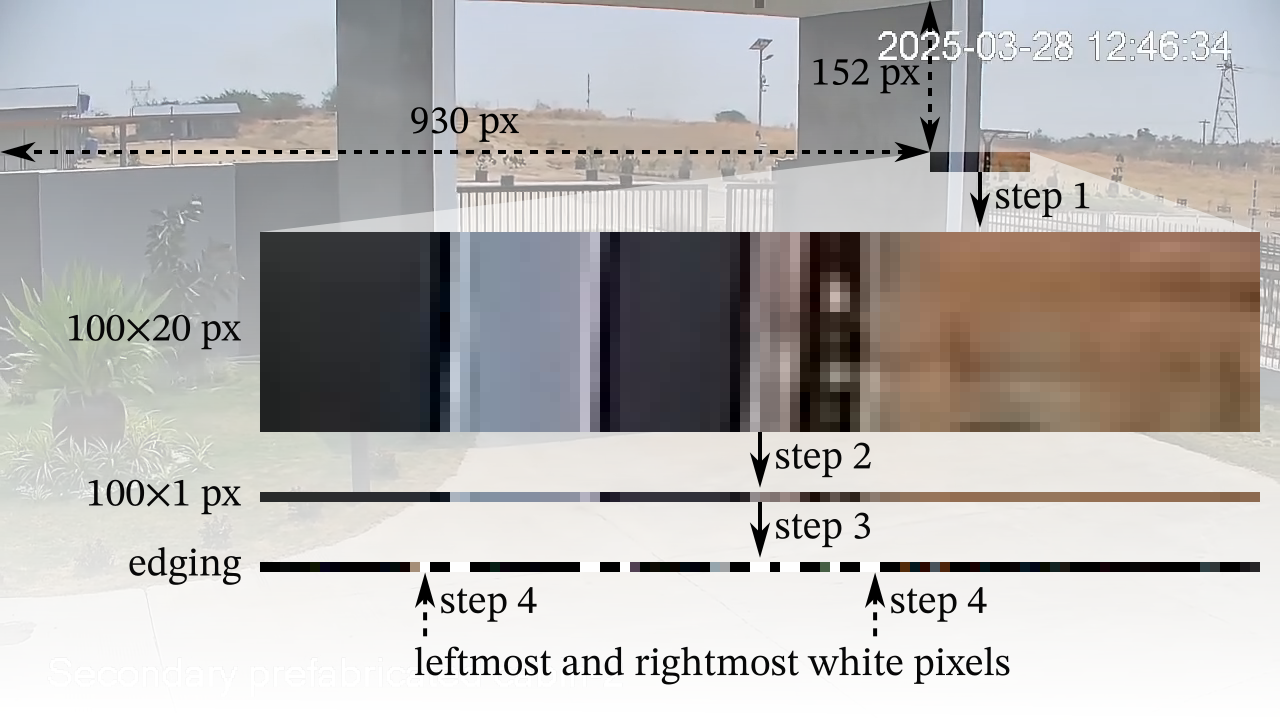}
\caption{Steps for image analysis to quantify slip history on the surface fault. The leftmost white pixel represents a part of the gate structure in front of the fault, while the rightmost one, a part of a pole on the opposite side.}
\label{VideoAnalysis}
\end{figure}

\begin{figure}[h!]
\centering
\includegraphics[width=\linewidth]{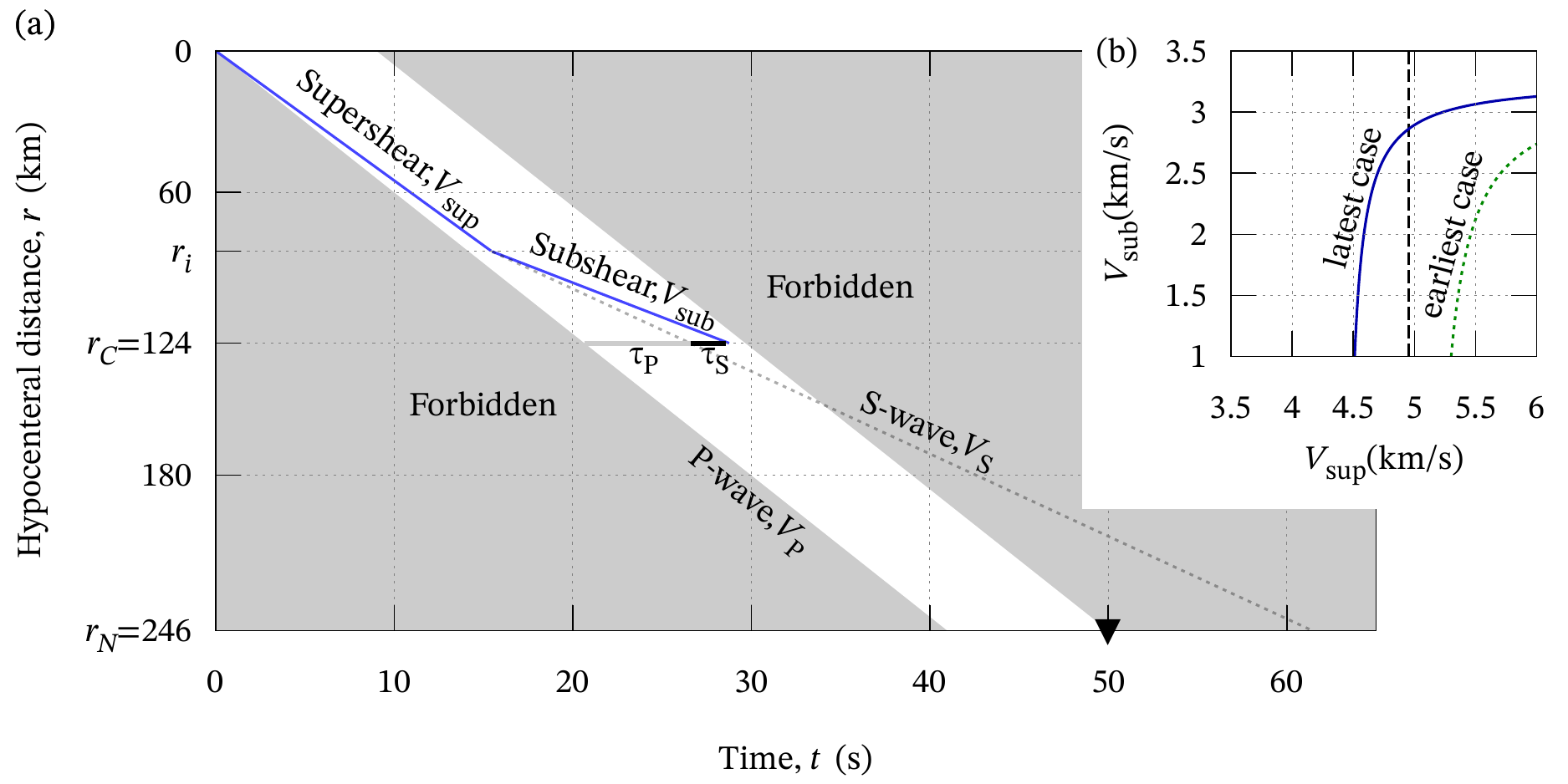}
\caption{(a) Schematic illustration of possible rupture front to CCTV $(r=r_C)$. The rupture front, with a velocity of up to $V_P$ from the origin to the black triangle (NPW), is unable to penetrate the gray-filled regions. At CCTV, the gray $(\tau_P)$ and the black $(\tau_S)$ horizontal bars correspond to the durations of the P-wave and S-wave, respectively. The deceleration point of the rupture front, $r_d$, is the intersection of the early supershear rupture front and the S-wave.
(b) Subshear rupture velocity as a function of the supershear rupture velocity, $V_\text{sup}$. The earliest case corresponds to $\tau_P = 3$ s, while the end of $\tau_S$ touches the forbidden region in the latest case.
The vertical dashed line in (b) represents $\sqrt{2} V_S$ (see the main text for its meaning).
}
\label{Algebra}
\end{figure}

\begin{figure}[h!]
\centering
\includegraphics[width=\linewidth]{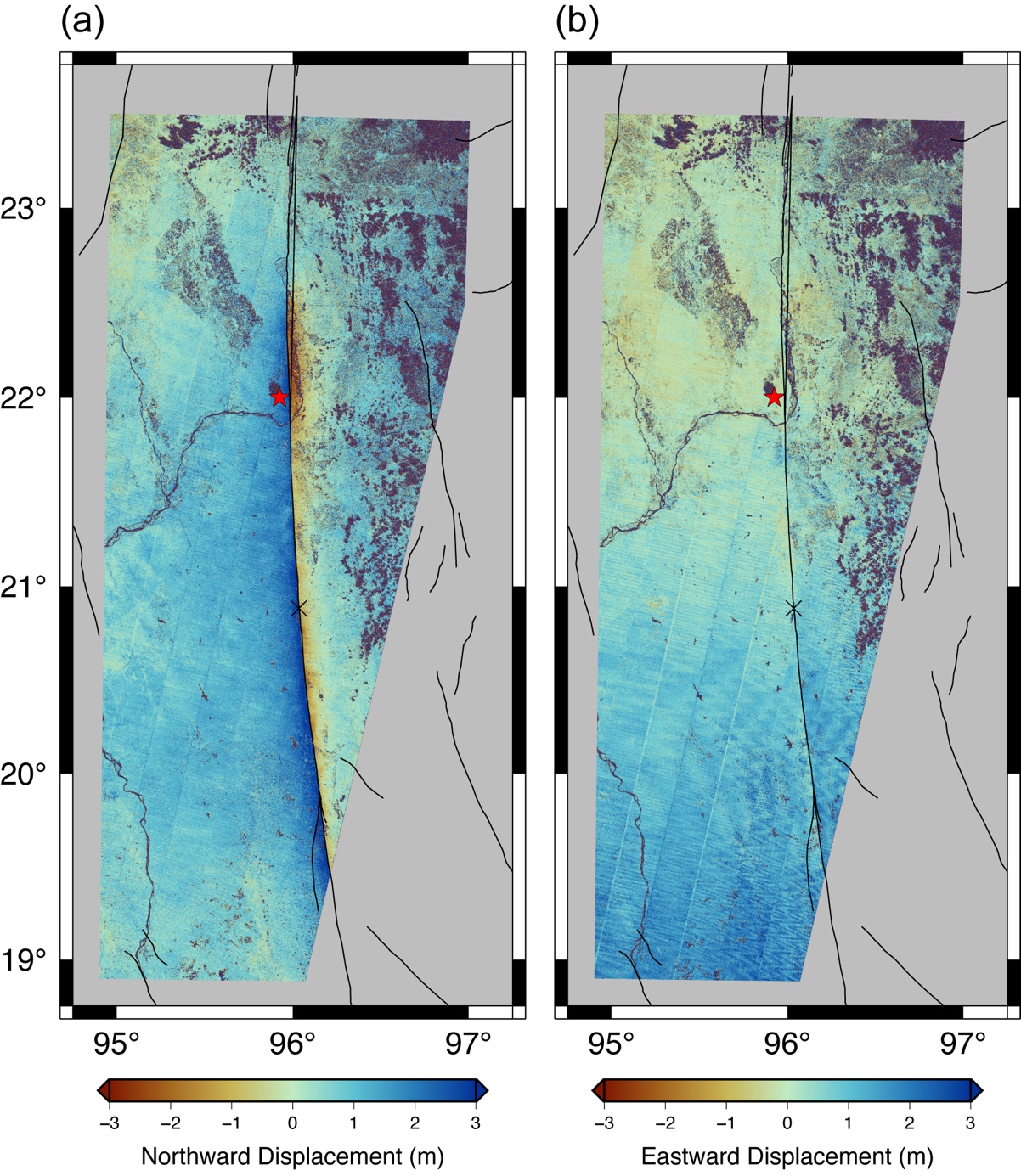}
\caption{Surface displacement estimated by the pixel-offset (optical correlation) analysis of Sentinel-2 imagery. (a) Northward and (b) Eastward displacements. The red star and the $\times$ mark show the locations of the epicenter and CCTV, respectively. Fault traces are based on \cite{Styron20}.}
\label{Sentinel2}
\end{figure}

\end{document}